\documentclass[preprint,showpacs,preprintnumbers,amsmath,amssymb]{revtex4}
\usepackage{graphicx}
\usepackage{epstopdf}
\usepackage{dcolumn}
\usepackage{bm}
\usepackage{hyperref}

\begin{document}
\preprint{}
\title{AdS Chern-Simons Gravity induces Conformal Gravity}

\author{Rodrigo Aros}
\email{raros@unab.cl}
\affiliation{Departamento de Ciencias Fisicas, Universidad Andres Bello, Av. Republica 252, Santiago,Chile}

\author{Danilo E. D\'{\i}az}
\email{danilodiaz@unab.cl}
\affiliation{Departamento de Ciencias Fisicas, Universidad Andres Bello, Av. Republica 252, Santiago,Chile}

\pacs{04.50.-h, 11.15.Yc, 04.50.Cd}
\begin{abstract}

\emph{The leitmotif of this paper is the question of whether four- and higher even-dimensional Conformal Gravities do have a Chern-Simons pedigree. We show that Weyl gravity can be obtained as dimensional reduction of a five-dimensional Chern-Simons action for a suitable (gauged-fixed, tractor-like) five-dimensional AdS connection. The gauge-fixing and dimensional reduction program admits a readily generalization to higher dimensions for the case of certain conformal gravities obtained by contractions of the Weyl tensor.}

\end{abstract}

\maketitle

\section{Introduction}

Four-dimensional conformal (Weyl) gravity has received a renewed interest since the advent of AdS/CFT correspondence.
In the interplay between the latter and conformal geometry, the Weyl action turns up in the form of certain conformally invariant terms in the volume renormalization of conformally compact Einstein (CCE) metrics~\cite{Graham:1999jg}:
\begin{enumerate}
  \item in `Lanczos-form', as volume anomaly of 5D  CCE metrics
\begin{equation}\label{Lanczos}
\int_{\mathcal{M}_4}\left(\mbox{Ric}^{\,2}-\frac{1}{3}\mbox{R}^{\,2}\right)
\end{equation}
given by the boundary integral of the so-called Q-curvature~\cite{BransonFirstQCurvatureApp1991} in 4D;
  \item in `Weyl-form', as renormalized volume of 4D CCE metrics~\cite{Anderson,Miskovic:2009bm}
\begin{equation}\label{Weyl}
\int_{\mathcal{M}_4} \mbox{Weyl}^{\,2}~.
\end{equation}
\end{enumerate}
The integral of Branson's Q-curvature generalizes the volume anomaly to higher even dimensions~\cite{GZ}, with the Fefferman-Graham obstruction tensor~\cite{FG85} for its metric variation generalizing the 4D Bach tensor, as shown in~\cite{GH04}. The renormalized volume of even-dimensional CCE metrics admits as well a higher-dimensional extension~\cite{CQY}. Both constructions result in particular candidates for conformal gravities in higher even dimensions. In six dimensions, for example, a particular combination of Weyl contractions has been singled out by the requirement that its space of solutions contains all Einstein metrics~\cite{Pope,Pope2}; one could have anticipated this result by recalling that one of the features of the Fefferman-Graham obstruction tensor is that it vanishes for conformally Einstein metrics, so that the resulting combination of Weyl terms is precisely the one in $Q_6$ as computed, for example in~\cite{BFT}, within AdS/CFT correspondence.

The aim of this letter is to gain a new perspective on four- and higher even-dimensional conformal gravities by addressing the question of whether they do admit a Chern-Simons (CS) formulation. The answer to the analogous question in three dimensions has long been known: the lagrangian of 3d conformal gravity of Deser, Jackiw and Templeton~\cite{Deser:1981wh,Deser:1982vy} is precisely the CS lagrangian of the {\em tractor connection}~\cite{Thomas26,Thomas32,BEG94,Eastwood96}. This is actually what Horne and Witten showed~\cite{Horne:1988jf}, even before the name tractor was coined in conformal geometry~\footnote{In fact, the tractor connection first appeared in the physics literature in~\cite{Crisp,Kaku} as gauging of the conformal group under vanishing torsion and tracefree curvature constraints. Weyl's garvity turned up then as a gauge theory of the conformal group; generalizations to higher dimensions made use of compensating fields.}.

In three dimensions, conformal gravity is constructed out of a dreibein $e_{\mu}^{\hspace{1ex} i}$ as fundamental variable and the  action\cite{Deser:1981wh,Deser:1982vy}
\begin{equation}\label{threeDAP}
I_{CG}=\int_{\mathcal{M}_3}  w_{i}\wedge d w^{i} +\frac{2}{3}\varepsilon^{ijk} w_{i}\wedge w_{j} \wedge w_{k}
\end{equation}
where $w_{i} = \varepsilon_{ijk}\, \omega_{\mu}^{\hspace{1ex}kl}\, dx^{\mu}$ with $\omega_{\mu}^{\hspace{1ex}kl}$ the Levi-Civita or Riemannian connection associated with the given dreibein $e^i$ so that, despite resemblance, this is not a Yang-Mills gauge theory. The (covariant) equation of motion demands three-dimensional spacetime to be conformally flat, \emph{i.e.}, a vanishing Cotton tensor
\begin{equation}
C_{\mu\nu\lambda} = \nabla_\mu \rho_{\nu\lambda}-\nabla_\lambda \rho_{\nu\mu}=0,
\end{equation}
the covariant curl of the Schouten or \emph{rho}-tensor
\begin{equation}
\rho_{\mu\nu}=R_{\mu\nu}-\frac{1}{4}R g_{\mu\nu}~.
\end{equation}

The tractor connection, on the other hand, comes into play from the conformal group in three dimensions. There are 10 generators: 3 translations ($P_i$), 3 Lorentz boosts and rotation ($J_{ij}$,  or alternatively `dualized' to a 3-vector $J_i$), 3 special conformal transformations ($K_i$)
and 1 dilatation ($D$). The gauge connection is the Lie-algebra valued form
\begin{equation}
A=e^{\;i}P_i+w^{\;i}J_i+\lambda^{\;i}K_i+\phi D~,
\end{equation}
and the Chern-Simons action for this gauge theory of $SO(3,2)$
\begin{equation}\label{ChernSimons3d}
I_{CS}=\frac{k}{8\pi}\int_{\mathcal{M}_3} \left\langle A \wedge dA + \frac{2}{3} A \wedge A \wedge A\right\rangle
\end{equation}
yields a vanishing curvature (or flat connection) as equation of motion
\begin{equation}
F=dA + A\wedge A = 0~.
\end{equation}

The classical equivalence between three-dimensional conformal gravity and this CS theory was established by Horne and Witten~\cite{Horne:1988jf}. Under the assumption of invertibility of the dreibein, they showed that the gauge choice $\phi_{\mu}=0$ is consistent and that the equations of motions, after drastic simplification, force $\omega_{\mu}^{\;ij}$ to be the Levi-Civita connection and $-\lambda_{\mu}^{\;i}$ to be the \emph{rho}-tensor. In all,
\begin{eqnarray}
  \phi &=& 0~,\\
  de^{i} + \omega^{i}_{\;j}\, e^{j} &=& 0~,\\
   -\lambda^i &=& \rho^i~,\\
  d\rho^{i} + \omega^{i}_{\;j}\,\rho^{j} &=& 0~.
\end{eqnarray}
The first three relations above define the gauge choice, whereas the last one is precisely the equation of motion of 3D conformal gravity: vanishing of the Cotton tensor.
In this particular gauge, the Chern-Simons action becomes that of conformal gravity, that is, the CS lagrangian of the resulting partially on-shell gauge-fixed connection
\begin{equation}
^{\tau}A_{\mu}=e_{\mu}^{\;i}P_i+w_{\mu}^{\;i}J_i-\rho_{\mu}^{\;i}K_i~.
\end{equation}
In component form, one can easily recognize the tractor connection of conformal geometry~\cite{BEG94,Eastwood96} (different conventions demand a little scrambling and sign flips)
\vspace{4mm}
\begin{center}
\begin{tabular}{c|c|c}
  0 & $-e_{\mu}^{\;i}$ & 0 \\\hline
  $e_{\mu}^{\;j}$ & $\omega_{\mu}^{\;ij}$ & $\rho_{\mu}^{\;j}$ \\\hline
  0 & $-\rho_{\mu}^{\;i}$ & 0
\end{tabular}
\;~.\end{center}
\vspace{4mm}

We do not dwell any further in odd dimensions as final target space; our present interest, instead, focuses in conformal gravities in four and higher even dimensions, where no direct construction via a CS form is available. A possibility, inspired by AdS/CFT correspondence, suggests itself: to look for a CS form in one dimension higher and trade conformal symmetry in $d=even$ by AdS group in $d+1=odd$. However, unlike usual AdS/CFT lore, contact with final even-dimensional target space will be achieved by dimensional reduction on a circle.

Our proposal is to start with AdS-invariant Chern-Simons lagrangians in odd dimensions and then to perform a suitable gauge-fixing that, after dimensional reduction, leads to local curvature invariant  lagrangians as candidates for conformal gravities in even dimensions. This new perspective on four- and higher even-dimensional conformal gravities may cast a different light on the problems on unitarity and renormalizability of gravitational theories (cf.~\cite{Pope, Maldacena:2011mk} for a recent discussion).
\section{AdS Chern-Simons Gravity}
Our starting point will be the theory of gravity in $d=2n+1$ obtained as Chern-Simons gauge theory for the $SO(2n,2$) group~\cite{Chamseddine:1989nu}
The writing in terms of spin connection for the Lorentz group proceeds as follows
\begin{enumerate}
   \item splitting of the general AdS-connection~\footnote{In a similar way, the extension of the Lorentz symmetry of the tangent space to $SO(2n+1,1)$ (de Sitter) or $ISO(2n,1$) (Poincar\'e) can be made.}
\begin{equation}\label{TheConnection}
A = \frac{1}{2}\hat{\omega}^{IJ} J_{IJ} = \frac{1}{2}\hat{\omega}^{ij} J_{ij} + q^i J_{i,2n+2},
\end{equation}
where $i,j=1,2,\ldots,2n+1$. More graphically,
\begin{center}
\begin{tabular}{c|c}
  $\omega_{\mu}^{\;ij}$ & \,$q_{\mu}^{\;j}$ \\\hline
  $-q_{\mu}^{\;i}$ & 0
\end{tabular}
~.\end{center}
\vspace{4mm}
   \item next, identifying $q^i$ and $\hat{\omega}^{ij}$  with the vielbein $e^i$ and the Lorentz spin-connection $\omega^{ij}$, respectively, on the manifold to be considered.
 \end{enumerate}

In this way, the AdS-connection is rewritten as
\begin{equation}\label{connection}
    A = \frac{1}{2}\omega^{ij} J_{ij} + e^i J_i~,
\end{equation}
where, for simplicity, we have set the AdS radius to one and renamed $J_{i,2n+2}\equiv J_i$ as a \emph{momentum} generator (which is not incorrect, but one has to keep in mind that $[J_{i},J_{j}] = J_{ij}$)~\footnote{It is worth mentioning that this splitting is not free of ambiguities. For instance: $A=0$ is a perfectly sound flat connection, but  would lead to $e^i=0$ which is not a sound vielbein for any manifold.}.

Little has been said about the r\^ole of the trace $\langle\rangle$ in the algebra; nonetheless, for any even dimension $d=2n$, the following trace (invariant tensor) will be used
\begin{equation}
\langle J_{I_1 I_2} \ldots J_{I_{2n+1} I_{2n+2}}\rangle = \varepsilon_{I_{1}\ldots I_{2n+2}},
\end{equation}
which, throughout $I=(i,2n+2)$ with $i=1\ldots 2n+1$, amounts to the trace to be considered hereafter
\begin{equation}
\langle J_{i_1 i_2} \ldots J_{i_{2n-1} i_{2n}} J_{i_{2n+1}}\rangle = \varepsilon_{i_{1} \ldots i_{2n+1}}.
\end{equation}

We close this section with a final remark on AdS Chern-Simons gravity. The action, modulo boundary terms, can be rewritten in the form of a (first order) Lovelock gravity~\cite{Lovelock:1971yv,Zanelli:2002qm} as
\begin{widetext}
\begin{equation}\label{CSinGeneralLovelock}
I_{CS} = \int \sum_{p=0}^{n} \frac{1}{2n+1-2p}\left(\begin{array}{c}
                                                    n \\
                                                    p
                                                  \end{array}\right)\varepsilon_{i_{1}\ldots i_{2n+1}} {R}^{i_{1} i_{2}}\ldots
                                                  {R}^{i_{2p-1} i_{2p}} e^{i_{2p+1}}\ldots e^{i_{2n+1}}
\end{equation}
\end{widetext}
where ${R}^{ij} = d{\omega}^{ij} + {\omega}^{i}_{\hspace{1ex} k}\, {\omega}^{kj}
$ with $i,j,k=1,\ldots ,2n+1$. It is worthwhile to recall that the vielbein is merely a part of the connection,\emph{ e.g.}, $e^i = \omega^{i,2n+1}$. The equations of motion of Chern-Simons gravity are generically
\begin{eqnarray*}
 \langle\delta A F^n\rangle &=& \delta \hat{\omega}^{I_{1} I_{2}} F^{I_{3} I_{4}}\ldots F^{I_{2n+1} I_{2n+2}} \varepsilon_{I_{1}\ldots I_{2n+2}} = 0  \\
 &=& \delta e^{i_{2n+1}} \varepsilon_{i_{1}\ldots i_{2n+1}} \bar{R}^{i_{1} i_{2}}\ldots \bar{R}^{i_{2n-1} i_{2n}}\\
 &+& \delta \omega^{i_{2n} i_{2n+1}} \varepsilon_{i_{1}\ldots i_{2n+1}} \bar{R}^{i_{1} i_{2}}\ldots \bar{R}^{i_{2n-3} i_{2n-2}} T^{i_{2n-1}}=0.
\end{eqnarray*}
with $\bar{R}$ stands for $\bar{R}^{ij}= R^{ij} +  e^i\,e^j$ and $T^i = de^i + \omega^{i}_{\hspace{1ex} j} e^j$ stands for torsion two-form. For simplicity, and in order to connect with Lovelock equation of motion, this is usually split as
\begin{eqnarray*}
  \mathcal{E}_{i_{2n+1}}=\varepsilon_{i_{1}\ldots i_{2n+1}} \bar{R}^{i_{1} i_{2}}\ldots \bar{R}^{i_{2n-1} i_{2n}} &=& 0 \\
   \mathcal{E}_{i_{2n} i_{2n+1}}=\varepsilon_{i_{1}\ldots i_{2n+1}} \bar{R}^{i_{1} i_{2}}\ldots \bar{R}^{i_{2n-3} i_{2n-2}} T^{i_{2n-1}}&=& 0
\end{eqnarray*}

It must be stressed that, unlike for any other Lovelock gravity, to impose $T^{i}=0$ is actually not the most general solution in this case. This is due the fact that the torsion two-form in the context of Chern-Simons gauge theory is merely a component of the gauge curvature, $T^i = F^{i 2n+1}$, and therefore $T^i=0$ comprises a very particular subset of the space of solutions of $F^n=0$.

\section{A tractor-like gauging of the AdS connection}
 The idea now is to reformulate a conformal theory of gravity in even dimensions as a Chern-Simons gauge theory with the help of an extension of a manifestly conformally invariant tractor-like connection. The construction is obviously not direct because of the clash between the numbers of dimensions: there is a tractor connection for SO($d-1,2$) in the (even) $d-1$-dimensional manifold, whereas the SO($d-1,2$) CS density lives in (odd) $d$ dimensions. The way out proposed in this work is to proceed with a dimensional reduction of the $2n+1$-CS density to end up with a $2n$-dimensional manifestly conformally invariant theory. The simplest approach assumes a $2n+1$-manifold of the form $\mathcal{M}' = \mathcal{M}\times S^{1}$ or $\mathcal{M}' = \mathcal{M}\times \mathbb{R}^{1}$ and considers only the zero modes, so that the difference in this case becomes irrelevant.

In close analogy with a tractor connection~\cite{Eastwood96}, we consider a connection for the SO($2n,2$) group in terms of the conformal generators (see Appendix \ref{Appendix-A} for their expression) in the space $\mathcal{M}'=\mathcal{M} \times S^{1}$ given by
\begin{equation}\label{AdS-tractor}
A = \frac{1}{2}\omega^{ij}(x)J_{ij} + e^{i}(x)P_{i}+\rho^{i}(x) K_{i} + \Phi(x) d\varphi D
\end{equation}
where $i,j=1,2,\ldots 2n$ and a system of coordinates  $X^{M}=(x^{\mu},\varphi)$ has been considered on $\mathcal{M}'$ with $\varphi$ parametrizing $S^{1}$.
On the other hand,
\[
 \rho^{i} =  e^{i}_{\hspace{1ex}\nu} \rho^{\nu}_{\hspace{1ex}\mu} dx^{\mu}
\]
with $\rho^{\mu}_{\hspace{1ex}\nu}$
\begin{equation}\label{SchoutenTensor}
\rho^{\nu}_{\hspace{1ex}\mu}  = \frac{1}{d-3} \left( R^{\nu}_{\hspace{1ex}\mu} - \frac{1}{2(d-2)}  \delta^{\nu}_{\mu} R \right)  \end{equation}
being the Schouten tensor of the $d-1=2n$-dimensional $\mathcal{M}$ and $R^{\nu}_{\hspace{1ex}\mu}$ and $R$, the Ricci tensor and scalar, respectively. The Schouten tensor relates the Riemann and Weyl curvatures
\[
R_{\mu\nu\alpha\beta}=W_{\mu\nu\alpha\beta} + g_{\mu\alpha} \rho_{\nu\beta}-g_{\nu\alpha} \rho_{\mu\beta}-g_{\mu\alpha} \rho_{\nu\alpha}+g_{\nu\beta} \rho_{\mu\alpha}~,
\]
or, equivalently, as two-forms
\begin{equation}\label{TheMostImportantRelation}
R^{ij} = \frac{1}{2} W^{ij}_{\hspace{2ex} k l} e^{k} e^{l} + 2(e^{i} \rho^{j} - e^{j}\rho^{i})
\end{equation}
where $W^{ij}_{\hspace{2ex} k l}$ is the Weyl tensor.

\subsection*{Weyl Transformation}

This tractor-like connection is constructed to make explicit a Weyl transformation on $\mathcal{M}$ in the form of a gauge transformation $A \rightarrow e^{\xi D} A e^{-\xi D} + e^{\xi D} d( e^{-\xi D})$. In this case the transformation of the component of $A$ in Eq.(\ref{AdS-tractor}) transforms as
\begin{eqnarray*}
  e^{i} &\rightarrow & e^{\xi} e^{i} \\
  \omega^{ij}  &\rightarrow & \omega^{ij} + \Upsilon^{i} e^{j} -  \Upsilon^{j} e^{i}  \\
  \rho^{i} &\rightarrow& e^{-\xi}(\rho^{i} + D \Upsilon^{i} + \Upsilon^{i}\Upsilon_{\mu} dx^{\mu} + e^{i} \Upsilon_{\mu} \Upsilon^{\mu} )
\end{eqnarray*}
with $\Upsilon_{\mu} = \partial_{\mu} \xi(x)$ and $\Upsilon^{i} = E^{i\mu}\Upsilon_{\mu} = E^{i\mu}\partial_{\mu} \xi(x)$.

The introduction of $\Phi$ along $D$ does not changes the law of transformation of the other component of the connection under $A \rightarrow e^{\xi D} A e^{-\xi D} + e^{\xi D} d( e^{-\xi D})$. In fact this only introduces a transformation for $\Phi$ given by
\[
\Phi d\varphi \rightarrow \Phi d\varphi - d\xi
\]
where $d\xi$ has only projection on $\mathcal{M}$ which determine in practice that a Weyl transformation, from the point of view of its pullback on $\mathcal{M}$ has no effect on $\Phi$.  Furthermore this transformation, as it will be observed, has no effect at all on the CS action. This defines that $\Phi$, from the point of view of the effective theory under discussion, is actually a scalar field under Weyl transformation.

\section{Dimensional reduction: from 5 to 4 dimensions}

Let us illustrate the dimensional reduction from 5 to 4 dimensions. In this case one starts with a product manifold $\mathcal{M}' = \mathcal{M}\times S^{1}$, coordinates $X^{M}=(x^{\mu},\varphi)$, and a gauge-fixed connection the form
\begin{equation}\label{TractorConnectionImproved}
    A =  \frac{1}{2}\omega^{ij} J_{ij} + (e^{i}+\rho^i)J_{i6} + (e^{i}-\rho^{i}) J_{i 5} + \Phi d\varphi D
\end{equation}
where $i=1 \ldots 4$.

For $d=5$ and AdS$_{5}$ (SO(4,2)) the idea relays on considering the 5-Chern Simons density
\begin{equation}\label{CSdensity}
 I = \int_{\mathcal{M}_4\times S^{1}} \varepsilon_{abcdf}\left(\hat{R}^{ab} \hat{R}^{cd} q^{f} + \frac{2}{3} \hat{R}^{ab} q^{c} q^{d} q^{f} + \frac{1}{5} q^{a} q^{b} q^{c} q^{d} q^{f}\right),
\end{equation}
where $\hat{R}^{ab} = d\hat{\omega}^{ab} + \hat{\omega}^{a}_{\hspace{1ex} c} \hat{\omega}^{cb}
$ with $a=1,\ldots ,5$, and establishing the map between $e^{i},\rho^i$ and $q^{a}$ as follows
\begin{eqnarray}
  \hat{\omega}^{i j}  &=& \omega^{ij}\nonumber\\
  \hat{\omega}^{i\,5} &=&e^{i} - \rho^i \nonumber \\
  \hat{\omega}^{5\,6} &=& \Phi(x)d\varphi = q^5\label{FourdimensionalExpasionofConnections}\\
  \hat{\omega}^{i\,6} &=& e^{i}+\rho^{i} = q^i \nonumber~.
\end{eqnarray}
This yields
\begin{eqnarray}
  \hat{R}^{ij} &=& R^{ij}-(e^{i} - \rho^{i})(e^{j}-\rho^{j}) \textrm{ and } \nonumber\\
  \hat{R}^{i5} &=&  D(e^{i}+\rho^{i})\label{FourdimensionalExpasionofCurvatures},
\end{eqnarray}
where now $R^{ij} = d\omega^{ij} + \omega^{i}_{\hspace{1ex} k}\omega^{k j}$ is the four-dimensional Riemann curvature two-form.

After replacing Eq.(\ref{FourdimensionalExpasionofCurvatures}) and Eq.(\ref{FourdimensionalExpasionofConnections}) into Eq.(\ref{CSdensity}) and integrating along $\varphi$ one obtains the four dimensional action
\begin{equation}\label{LagrangianReduced}
I = \int_{\mathcal{M}_4} \Phi\varepsilon_{ijkl}\left( R^{ij}R^{kl} + 8 R^{ij}e^{k} \rho^{l} + 16  e^{i} e^{j} \rho^{k} \rho^{l}\right)
\end{equation}

For the sake of clarity these terms can be rewritten in metric formalism. For instance,
\begin{equation}\label{EulerTermDensity}
 \varepsilon_{ijkl}R^{ij}R^{kl} =\left(R^{\mu\nu\alpha\beta}R_{\mu\nu\alpha\beta}-4 R_{\mu\nu} R^{\mu\nu}+R^2\right)\sqrt{g}d^{4}x =\mathcal{E}\sqrt{g}d^{4}x
\end{equation}
which can be recognized as the Euler density $\mathcal{E}$.
This defines, once expressed in metric formalism, the action
\begin{eqnarray}
  I  &=&  \int_{\mathcal{M}_4}\Phi \left[\mathcal{E} + 2 \left(\mbox{Ric}^{\,2}-\frac{1}{3}\mbox{R}^{\,2}\right)\right]\label{LagrangianReducedCoor}\\
      &=&  \int_{\mathcal{M}_4} \Phi\cdot \mbox{Weyl}^{\,2}~.
\end{eqnarray}
The equations of motion for this action can be obtained either from the Chern-Simons equations of motion and later replacing the gauge choice, or directly from the variation of Eq.(\ref{LagrangianReducedCoor}) with respect to the metric. The result in both cases is a generalization of the Bach tensor. Moreover, a further gauging of $\Phi$ to a constant reduces the equation of motion to those of Weyl gravity, containing in particular all Einstein metrics.

\section{Dimensional reduction in higher dimensions}

The $2n+1$ dimensional Chern Simons action for SO($2n$,2) can be written as
\begin{widetext}
\begin{equation}\label{CSinGeneral}
I^{2n+1}_{CS} = \int_{\mathcal{M}_{2n}\times S^{1}} \sum_{p=0}^{n} \frac{1}{2n+1-2p}\left(\begin{array}{c}
                                                    n \\
                                                    p
                                                  \end{array}\right)\varepsilon_{a_{1}\ldots a_{2n+1}} \hat{R}^{a_{1} a_{2}}\ldots
                                                  \hat{R}^{a_{2p-1} a_{2p}} q^{a_{2p+1}}\ldots q^{2n+1}
\end{equation}
\end{widetext}
where $\hat{R}^{ab} = d\hat{\omega}^{ab} + \hat{\omega}^{a}_{\hspace{1ex} c}\, \hat{\omega}^{cb}
$ with $a,b,c=1,\ldots ,2n+1$. It is direct to prove after that replacing the ansatz
\begin{eqnarray}
  \hat{\omega}^{i j}  &=& \omega^{ij}\nonumber\\
  \hat{\omega}^{i\,2n+1} &=& e^{i} + \rho^i \nonumber \\
  \hat{\omega}^{2n+1\,2n+2} &=& \Phi(x)d\varphi = q^{2n+1}\label{2N+1dimensionalExpasionofConnections}\\
  \hat{\omega}^{i\,2n+2} &=& e^{i} - \rho^i = q^i, \nonumber
\end{eqnarray}
with $i=1,\ldots,2n$, the action merely becomes
\begin{eqnarray}
  I_{CS}^{2n+1} &=& \int_{\mathcal{M}_{2n}\times S^{1}} \varepsilon_{i_{1}\ldots i_{2n}}\left((R^{i_{1} i_{2}} + 2\rho^{i_{1}} e^{i_2}-2\rho^{i_{2}} e^{i_1})\ldots \right.\nonumber\\
   &\ldots& \left.(R^{i_{2n-1} i_{2n}} + 2\rho^{i_{2n-1}} e^{i_{2n}}-2\rho^{i_{2n}} e^{i_{2n-1}}) \right)\Phi d\varphi \label{LagrangianReduced2n}
\end{eqnarray}

This is direct to integrate along $\varphi$ and after some realignments in terms of Eq.(\ref{TheMostImportantRelation}), namely $R^{ij} = \frac{1}{2} W^{ij}_{\hspace{2ex} k l} e^{k} e^{l} - 2(e^{i} \rho^{j} - e^{j}\rho^{i})$, one can rewrite Eq.(\ref{LagrangianReduced2n}) as
\begin{equation}\label{LagrangianReduced2nWeyl}
I_{CG}^{2n} = \int_{\mathcal{M}_{2n}} \Phi \delta_{i_{1}\ldots i_{2n}}^{j_{1}\ldots j_{2n}}\left(W^{i_{1} i_{2}}_{\hspace{2ex}j_{1}j_{2}}\ldots W^{i_{2n-1} i_{2n}}_{\hspace{6ex}j_{2n-1} j_{2n}}\right)  \sqrt{g}\, d^{2n} x.
\end{equation}

These gravities just constructed, as the number of dimensions increases, saturate every possible contraction of the Weyl tensor to the corresponding power.  However Eq.(\ref{LagrangianReduced2nWeyl}) contains more than a mere collection of the possible contraction of the Weyl tensor, they are added with particular set of relative coefficients. This is due to the larger, but hidden, AdS symmetry preserved by this action principle, beyond the pure Weyl invariance preserved by an arbitrary set of coefficients. This is completely analogous to the general Lovelock action which preserves Lorentz symmetry versus the Chern-Simons action, a Lovelock gravity with particular set of coefficients, which preserves the larger AdS symmetry.

Essentially Eq.(\ref{LagrangianReduced2nWeyl}) has the same form of the Euler density but in this case the Riemann curvature has been replaced by the Weyl curvature.
The seven dimensional case gives rise to a six-dimensional conformal gravity
\begin{equation}\label{ConformaGravity7}
I^{6}_{CG} = \int_{\mathcal{M}_6} \Phi \left(\,W^{\mu\nu\gamma\tau}W_{\gamma\tau\xi\lambda}W^{\xi\lambda}_{\hspace{2ex}\mu\nu} + 4\; W^{\mu\nu}_{\hspace{2ex}\xi\lambda}W^{\xi\tau}_{\hspace{2ex}\nu\kappa}W^{\lambda\kappa}_{\hspace{2ex}\mu\tau}\,\right)\sqrt{|g|} d^6 x~.
\end{equation}

\section{Conclusion and outlook}

In retrospective, in this work it has been shown that a family of conformal gravities in even dimensions, constructed out of the Weyl tensor and a scalar field (under diffeormorphisms and Weyl transformation), can be cast as a Chern Simons theory for the conformal or AdS group in the appropriate dimension. Furthermore, it has been shown that certain combinations of the Weyl invariants present an enlarged, and seemingly unnoticed,  SO($2n,2$) symmetry. The purely gravitational theories that result upon gauging of $\Phi$ to a constant ought to be compared with those obtained in~\cite{Kastor:2013uba} based on squares of higher curvature Weyl tensors.

It is worthwhile to stress that although the $2n$ dimensional theory, after compactification, is purely metric actually it is not possible to translate the original (gauge-fixed) Chern Simons theory into a purely metric $2n+1$ Lovelock theory. In $2n+1$ dimensions in addition to the \emph{translation} from vielbein into metric it is necessary the introduction of a non-Levi-Civita connection as the ansatz itself defines a non zero torsion in $2n+1$ dimensions.

Another open issue concerns the possibility to implement different traces, i.e., different invariant tensors, and the question of whether these exhaust the list of type-B trace anomalies in the given even dimension. This would require the bystander field $\Phi$ to be gauged to a constant; however, we notice that a kinetic term for this field would demand a (conformally invariant) differential operator of order equal to the dimensionality of the spacetime and the action would look as a higher dimensional version of the (local) Riegert's action~\cite{Riegert}. We also notice similarities with the approach in~\cite{Anabalon:2006fj}, although our tractor-like gauging eventually leads to Weyl rather than Einstein gravity.

In all, the formulation of conformal gravities as CS theories may have some interesting consequences and open up an alternative route to address unitarity and quantization aspects of gravitational theories. In principle, the methods presented here could also be extended to any group containing AdS as a subgroup; one could contemplate the possibility for supersymmetric~\cite{Cham90,Troncoso:1999pk} and/or higher-spin formulations.

\acknowledgments

R. Aros would like to thank Prof. J. Zanelli for inspiring discussions during the last 15 years, and also wish him a happy 60th birthday.
We also acknowledge useful conversations with O. Fierro, N. Merino, D. M. Pe\~nafiel, R. Olea and P. Sundell.
This work was partially funded by grants FONDECYT 1131075, 11110430 and UNAB DI-286-13/R, DI-295-13/R.

\appendix
\section{AdS and conformal algebras}
\label{Appendix-A}

Here we make explicit the connection between AdS$_{d}$ group and the conformal group in $(d-1)-$ dimensions for reference (see, e.g.,\cite{Aharony:1999ti}). On one hand the algebra of $AdS_{d}$ is given by
\begin{equation}\label{CommutationRelationsAdS}
\left[J_{AB},J_{CD}\right] = -\delta_{AB}^{EF} \delta_{CD}^{GH} \eta_{EG} J_{FH},
\end{equation}
with $A,B=0\dots d$, and on the other hand, the conformal algebra is
\begin{equation}\label{CommutationRelationsC}
\begin{array}{ll}
\left[M_{ij},M_{kl}\right] = -\delta_{ij}^{mn} \delta_{kl}^{op} \eta_{mo} M_{np}  &\\
\left[M_{ij},P_{k}\right] = -(\eta_{ik} P_{j}-\eta_{jk} P_{i})  &\left[D,P_{i}\right] = P_{i}  \\
\left[M_{ij},K_{k}\right] = - (\eta_{ik} K_{j}-\eta_{jk} K_{i}) &\left[D,K_{i}\right] = -K_{i} \\
\left[P_{i},K_{j}\right]= 2 M_{ij} + 2 \eta_{ij}D &  \left[D,M_{ij}\right] = 0 \\
\end{array}
\end{equation}
with $i,j=0\ldots d-2$. It is direct to observe that both sets map into each other throughout
\begin{equation}\label{CommutationRelationsAdSCMap}
\begin{array}{ll}
J_{ij} = M_{ij} & J_{id-1} = \frac{1}{2}(P_{i}+K_{i}) \\
J_{d-1 d} = D & J_{id} = \frac{1}{2}(P_{i}-K_{i}).
\end{array}
\end{equation}

\providecommand{\href}[2]{#2}\begingroup\raggedright\endgroup

\end{document}